%% file: np.tex
\documentclass[11pt,letterpaper]{article}
\usepackage[utf8]{inputenc}
\usepackage{amsmath}
\usepackage{enumitem}
\usepackage{amsfonts}
\usepackage{dcolumn}
\usepackage{amssymb}
\usepackage{subfig}
\usepackage{color}
\usepackage[hidelinks, colorlinks=true, citecolor=blue,
urlcolor=blue, linkcolor=blue]{hyperref}
\usepackage[sort]{natbib}
\usepackage{algorithm2e}
\usepackage{graphicx}
\graphicspath{{graphics/}}  
\usepackage{listings}
\usepackage{float}
\usepackage{booktabs}
\usepackage{setspace}
\usepackage{caption}
\usepackage{ntheorem}
\usepackage[left=1in,right=1in,top=2cm,bottom=2cm]{geometry}
\newtheorem{theorem}{Theorem}[section]

\newtheorem{assumption}{Assumption}
\newtheorem*{method*}{Procedure}

\author{Eliot Abrams\thanks{Booth School of Business, University of Chicago, \href{mailto:eabrams@uchicago.edu}{eabrams@uchicago.edu}} $\quad$ George Gui\thanks{University of Chicago, \href{mailto:georgegui@uchicago.edu}{georgegui@uchicago.edu}} $\quad$  Ali Horta\c{c}su\thanks{University of Chicago, \href{mailto:hortacsu@uchicago.edu}{hortacsu@uchicago.edu} \newline \indent We thank Patrick Bajari, Stéphane Bonhomme, Jean-Pierre Dubé, Jeremy Fox, Harikesh Nair, S. Sriram, and the participants of the AAAI-18: AI and Marketing Science Workshop and University of Chicago IO Workshop for invaluable conversations and comments. We also thank Alan Montgomery and Peter Rossi for providing background documentation on the experiments from \cite{hoch1994edlp}, the Kilts Center for Marketing for the Dominick's scanner data, and the Becker Friedman Institute IO Initiative for financial support.}}
\title{Finding Exogenous Variation in Data}
\begin{document}
\maketitle

\noindent \textit{\textbf{Abstract:} We reconsider the classic problem of recovering exogenous variation from an endogenous regressor. Two-stage least squares recovers exogenous variation through presuming the existence of an instrumental variable. We rely instead on the assumption that the regressor is a mixture of exogenous and endogenous observations--say as the result of temporary natural experiments. With this assumption, we propose an alternative two-stage method based on nonparametrically estimating a mixture model to recover a subset of the exogenous observations. We demonstrate that our method recovers exogenous observations in simulation and can be used to find pricing experiments hidden in grocery store scanner data.}
\vspace*{3px}
\vspace*{5px}

\doublespacing

\section{Introduction}
Consider the classic instrumental variables setup in which $Y = \alpha + \beta X + \epsilon$, $Cov(X, \epsilon) \ne 0$, and there exists a $Z$ such that $Cov(Z, \epsilon) = 0$ and $Cov(Z, X) \ne 0$. The problem faced here is that $X$ contains a mix of both exogenous variation that can be used to identify $\beta$ and endogenous variation that complicates this identification. The instrument $Z$ allows the researcher to isolate the exogenous variation. In the two-stage least squares (2SLS) solution, the researcher runs the first stage $X = \pi_0 + \pi_1 Z + u$ to recover the exogenous variation $\widetilde{X} = \pi_0 + \pi_1 Z$. The researcher then uses this recovered exogenous variation in a second stage to construct a consistent estimate for $\beta$.

The main roadblock to applying the 2SLS solution generally is finding a suitable $Z$. That said, the 2SLS approach conceptually applies quite broadly. As a prototypical example, consider a widget store choosing prices for its widgets. The researcher may see prices and quantities sold for each week and wish to recover consumers' price elasticity of demand. Unfortunately, in most weeks the widget store's profit maximizing prices are endogenous with the quantity sold thanks to the store's optimization problem. Fortunately, the widget store may also spend several weeks experimenting with its pricing strategy. If so, the observed sequence of widget prices contain both endogenous and exogenous variation.

Motivated by such settings, we revisit the problem of extracting exogenous variation with a new approach. We show that when $X$ is a mixture random variable with exogenous and endogenous components, we can use nonparametric estimation to recover a subset of the exogenous observations. These recovered exogenous observations can be used in a second stage to identify the parameters of interest just as with 2SLS.

We apply our approach to simulated data as a first proof of its effectiveness. We then use our method to try to find pricing experiments hidden in retail scanner data. Specifically, we consider the Dominick's scanner data maintained by the Kilts Center for Marketing (Dominick's is a former Chicago-area grocery store chain). \cite{hoch1994edlp} conducted a number of pricing experiments with Dominick's in the time period covered by this data. We demonstrate that our method recovers Hoch et al.'s ``Study 1" pricing experiments with high accuracy. In fact, we provide some evidence that our approach provides a more accurate description of the experiments than what remains of the original experiment documentation. 

Our work is an operationalization of the recent literature on nonparametric estimation of mixture models. \cite{hall2003nonparametric} consider a mixture of two component distributions each with $k$ independent coordinates and prove that nonparametric identification holds when $k\ge3$. \cite{benaglia2009like} present an expectation maximization-like algorithm for nonparametrically estimating a finite mixture of $m$ arbitrary component distributions with $r$ independent coordinates when $2^r-1 \ge mr+1$. \cite{bonhomme2016non} derive asymptotic results in the sub-case where the $r$ coordinates are independent and identically distributed.

We also parallel research on the estimation of mixture linear regressions. \cite{bashir2012robust} consider a model where there are $k$ latent populations each satisfying a linear model $Y_k = X_k^T \beta_k + \epsilon_k$. Assuming that the errors $\epsilon_k \sim N(0,\sigma_k)$ are independent across populations, the authors provide an expectation maximization algorithm to recover the class labels, $\beta_k$, and $\sigma_k$. Our setting differs from the above because we allow for a population $k\prime$ that has $X_{k\prime}$ endogenous with $\epsilon_{k\prime}$. Specifying the joint distribution of $X_{k\prime}$ and $\epsilon_{k\prime}$ would facilitate parametric identification in our setting following the authors' approach.

Finally, our work extends a computer science literature on using machine learning to identify causal relations. \cite{jensen2008automatic} propose a system for automatically identifying quasi-experimental designs from relational databases. \cite{grosse2016identification} develop an algorithm for inferring causal relations in brain imaging data. Most closely related, \cite{sharma2016split} consider settings where $Y$ can be split into (1) a random variable $Y_R$ that is caused by $X$ and (2) a random variable $Y_D$ that is independent of $X$ if there are no confounding unobserved variables that cause both $Y$ and $X$. Here, the authors can estimate the causal impact of $X$ on $Y_R$ from subsets of the data where $X \perp Y_D$. In contrast, our method applies for any $Y$ and instead proceeds by splitting $X$ into exogenous and endogenous observations.

The rest of the paper is organized as follows. Section \ref{method} presents our method. Section \ref{simulation} demonstrates our method on simulated data. Section \ref{application} takes our method to the Dominick's grocery store scanner data. Section \ref{conclusion} concludes. All figures and tables are in the respective appendices.

\section{Method}
\label{method}
Consider a random vector $(Y, X, \epsilon)$ satisfying $Y = \alpha + \beta X + \epsilon$ where $Cov(X, \epsilon) \ne 0$. 2SLS recovers exogenous variation from $X$ through assuming the existence of an instrument $Z$ satisfying $Cov(Z, \epsilon) = 0$ and $Cov(Z, X) \ne 0$. However, few applications feature an instrument that meets both conditions. As such, we propose a new method for recovering exogenous variation from $X$ in certain settings that does not rely on having a valid instrument.

We consider settings typified by the widget store example above wherein the store experiments with its prices in some weeks and not in others. Here the price of the widget in a week, $X$, is a mixture of two random variables. Let $f_1(x)$ be the density of the price in a non-experiment week and $f_2(x)$ be the density of the price in an experiment week. The density of $X$ is $f(x) = (1-P(\text{Experiment})) f_1(x) +  P(\text{Experiment}) f_2(x)$. Our approach is to estimate this mixture model, namely $P(\text{Experiment})$, $f_1(x)$, and $f_2(x)$. With these probabilities in hand, we can then find prices that are likely from the store's experiments--a set of observations of $X$ that are likely realized according to $f_2$.

However, without knowledge of which weeks are experiments, a researcher could not generally separately identify the component weights, i.e. $1-P(\text{Experiment})$ and $P(\text{Experiment})$, and the coordinate densities, $f_1(x)$ and $f_2(x)$. Fortunately, this is just a problem of dimensionality. When the mixture components have three or more independent coordinates, then it is possible to nonparametrically recover the component weights and the coordinate densities up to a re-labeling of the components. \cite{benaglia2009like} provide a consistent estimator.

In terms of the motivating example, assume the widget store sells three widgets with prices $X$, $W_1$, and $W_2$ that are independent of each other conditional on being in an experiment week or not. That is, 
\begin{align*}
f(x, w_1, w_2) &= (1-P(\text{Experiment})) f_{X,1}(x) f_{W_1,1}(w_1) f_{W_2,1}(w_2) \\& + P(\text{Experiment}) f_{X,2}(x) f_{W_1,2}(w_1) f_{W_2,2}(w_2)
\end{align*}
\noindent Then, the researcher can recover the two component weights and the six coordinate densities. The additional prices, $W_1$, and $W_2$, allow the researcher to disentangle the component weights from the coordinate densities. Importantly, $W_1$ and $W_2$ can be ``outside" variables that are not of direct interest to the researcher. The only restriction is that $W_1$ and $W_2$ constitute a mixture distribution with $X$. Also, it is important to repeat that this identification is only up to relabeling of the components. The estimation returns $\hat{f}_{X,a}(x), \hat{f}_{W_1,a}(w_1), \hat{f}_{W_2,a}(w_2)$ and separately $\hat{f}_{X,b}(x), \hat{f}_{W_1,b}(w_1), \hat{f}_{W_2,b}(w_2)$. The researcher needs to make an additional assumption to label component $a$ as the non-experiment component ($1$) and to label component $b$ as the experiment component ($2$).


To proceed, we then make Assumption \ref{one}:
\begin{assumption}
\label{one}
\noindent
\begin{enumerate}
\item $(Y, X, \epsilon, W_1, W_2)$ is a random vector satisfying $Y = \alpha + \beta X + \epsilon$
\item The sub-vector $(X, W_1, W_2)$ is a two component mixture model defined by the density function $f(x, w_1, w_2) = (1-\pi) f_{X,1}(x) f_{W_1,1}(w_1) f_{W_2,1}(w_2) + \pi f_{X,2}(x) f_{W_1,2}(w_1) f_{W_2,2}(w_2)$.\footnote{With $f_{X,1}(x) \ne f_{X,2}(x)$, $f_{W_1,1}(w_1) \ne f_{W_1,2}(w_1)$, and $f_{W_2,1}(w_2) \ne f_{W_2,2}(w_2)$.} 
\item For $X_1 \sim F_{X,1}(x)$ and $X_2 \sim F_{X,2}(x)$, $Cov(X_1, \epsilon)\ne0$ and $Cov(X_2, \epsilon)=0$
\item The researcher has $T$ iid observations on $(Y,X,W_1,W_2)$ and knows either (A) whether $p>0.5$ or (B) how a moment of a coordinate of $(X, W_1, W_2)$ differs between the two mixture components, e.g. $E(X_1) > E(X_2)$
\end{enumerate}
\end{assumption}

The first part of the assumption posits the standard linear framework. This specification is overly restrictive. More generally, $Y$ can be a nonlinear function of $X$, $W_1$, and $W_2$. The second part of the assumption states that $(X, W_1, W_2)$ is a two component mixture model with three independent coordinates whose densities are distinct between the two components. Again, this is a base case. Our approach immediately extends to mixture models with more components and more coordinates.\footnote{However, statisticians have yet to prove necessary and sufficient conditions for nonparametric identification of mixture models with more than two components. See the explanation in \cite{benaglia2009like}.} The third part of the assumption is that the first component produces the endogenous observations and the second component produces the exogenous observations with respect to the outcome of interest. Note that unlike 2SLS, we do not require that $W_1$ and $W_2$ have zero covariance with $\epsilon$. Finally, the fourth part of the assumption allows the researcher to  uniquely label the estimated components.


Given this assumption, our method for recovering exogenous observations is to:

\begin{method*}
\noindent
\begin{enumerate}[noitemsep]
\item Nonparametrically estimate the mixture model defining $(X, W_1, W_2)$ using the algorithm provided by \citep{benaglia2009like} or an alternative consistent estimator
\item Label the two components based on assuming (A) $\pi > (\le) \; 0.5$ or (B) a moment condition
\item Label observations of $X$ that are drawn from $X_2$ with probability greater than or equal to some threshold $p$ as exogenous
\end{enumerate}
\end{method*}

Step (1) returns a consistent estimate for the component weights and the coordinate densities. Step (2) labels one component as exogenous, i.e. the observations drawn from this component are experiments, and the other as endogenous. Finally, as the number of observations approaches infinity, Step (3) gives a subset of the set of observations of $(Y_t, X_t)$ where $X_t$ is drawn from is an experiment with probability greater than or equal to $p$ under the true mixture model (by the consistency of Step 1). Call this subset $\chi(p)$.


We now prove that the researcher can use the observations in $\chi(p)$ to consistently estimate $\beta$. Specifically, Theorem \ref{first} below proves that the ordinary least squares estimator for $\beta$ using the observations in $\chi(p)$ is consistent as $T \to \infty$ and $p \to 1$ under minor additional assumptions (proof in Appendix \ref{proof}). We demonstrate with a simulation and application below that accurate labels are achievable in practice.

\begin{theorem}
\label{first}
Let $\chi(p)$ be a subset of the set of observations of $(Y_t, X_t)$ where $X_t$ is drawn from $X_2$ with probability greater than or equal to a threshold $p$ under the true mixture model. Assume (1) $|\chi(p)|$ approaches infinity as $T$ approaches infinity and (2) the $Cov(X,\epsilon|X\in\chi(p))$ and $Var(X|X\in\chi(p))$ are finite for all $p\in[0,1]$.\footnote{A sufficient condition is that at least one of the coordinates has exogenous and endogenous component densities with the property that the coordinate's exogenous component density's support is not a subset of the endogenous component density's support.} Then
\begin{align*}
\hat{\beta}^{OLS}_{\chi(p)} & \underset{T\to\infty}{\overset{p}{\to}} \frac{Cov(Y,X|X\in\chi(p)) }{Var(X|X\in\chi(p))}\\
& \underset{p\to 1}{\to} \frac{Cov(Y_2,X_2|X_2 \in \chi(1))}{Var(X_2|X_2 \in \chi(1))} \\
& = \beta
\end{align*}
where $Y_2$ indicates that the data-generating process for $Y$ is in terms of $X_2$ only.
\newline \noindent Proof: In Appendix \ref{proof}
\end{theorem}

\section{Simulation}
\label{simulation}
A simulation provides an immediate test of our approach. Assume $X_1$, $W_{1,1}$, $W_{2,1}$ are each independent $U(0,1)$ random variables and $X_2$, $W_{1,2}$, $W_{2,2}$ are each independent $U(0,2)$ random variables. Let $(X, W_1, W_2) = (X_1, W_{1,1}, W_{2,1})$ with probability $1-\pi=0.4$ else $=(X_2, W_{1,2}, W_{2,2})$. $(X, W_1, W_2)$ is a mixture model with two components and three independent coordinates. 

Consider a researcher who wants to understand the causal relationship between the above $X$ and some outcome variable $Y$. The truth is that $Y = 2 X + \epsilon$. However, $\epsilon = X_1 + W_{1,1} + W_{2,1} + v$ where $v \sim U(0,1)$. All the coordinates of the first component are endogenous with $\epsilon$ (think of the widget stores prices during a non-experiment week). As described above, the role of $W_1$ and $W_2$ is to allow the researcher to separately identify the marginal density of $X$ in component 1 (i.e. the density of $X_1$) from the marginal density of $X$ in component 2 (namely, the density of $X_2$). More generally, $W_1$ and $W_2$ could be part of the generating process for $Y$.

Say that the researcher has $T=2,000$ iid observations on $(Y, X, W_1, W_2)$ with which to recover $\beta$. The researcher could try running ordinary least squares on the entire sample. However, for our realizations, the resulting point estimate is 2.12, which is significantly different from 2 at the 1\% level. See Column 1 of Table \ref{simulation_regs}. Note that 2SLS is of no help here because $W_1$ and $W_2$ are neither relevant nor exogenous instruments.

\begin{figure}[H]
\caption{The following density plots illustrate the recovery of the exogenous variation, $X_1$, in the simulation. Here we consider $(X, W_1, W_2)$ drawn from a random vector of three $U(0,1)$ random variables with probability $w=0.4$ and from a random vector of three $U(0,2)$ random variables else. Applying the nonparametric mixture model estimation from \cite{benaglia2009like} to 2,000 iid observations of $(X, W_1, W_2)$, returns the following coordinate densities for each component. The density of the coordinate in the first component is in red and the density of the coordinate in the second component is in green. Assessed visually, the nonparametric estimation works quite well.}
\label{density}
\centering
\subfloat{
  \includegraphics[scale=.8]{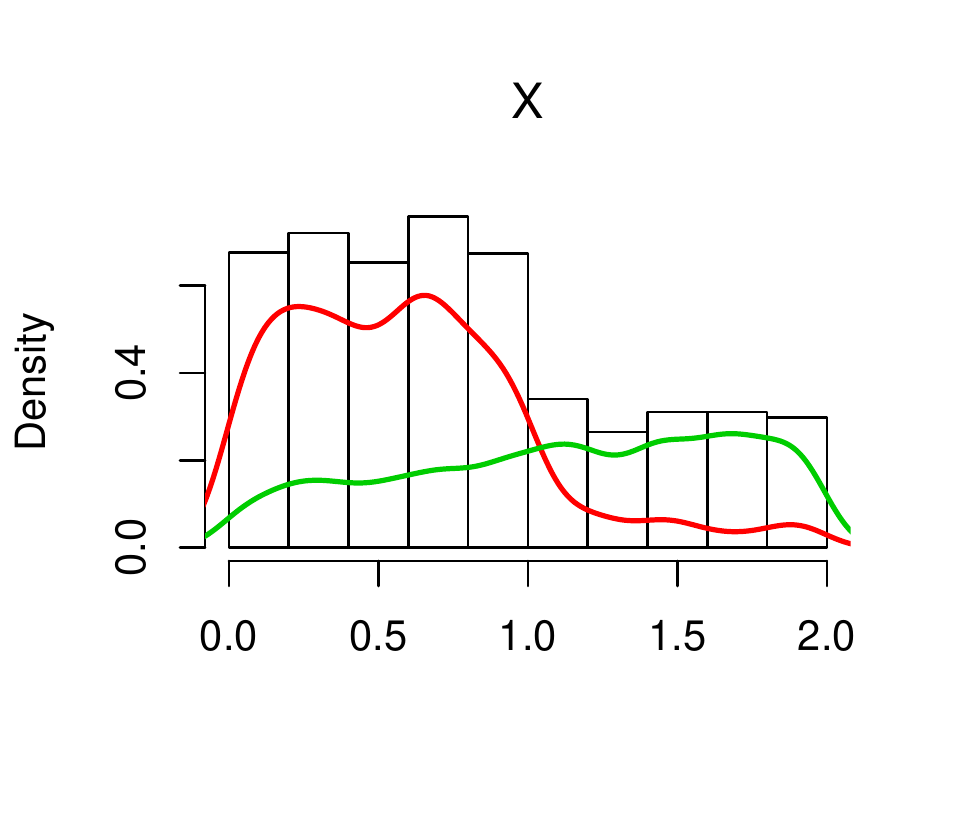}
}
\subfloat{
  \includegraphics[scale=.8]{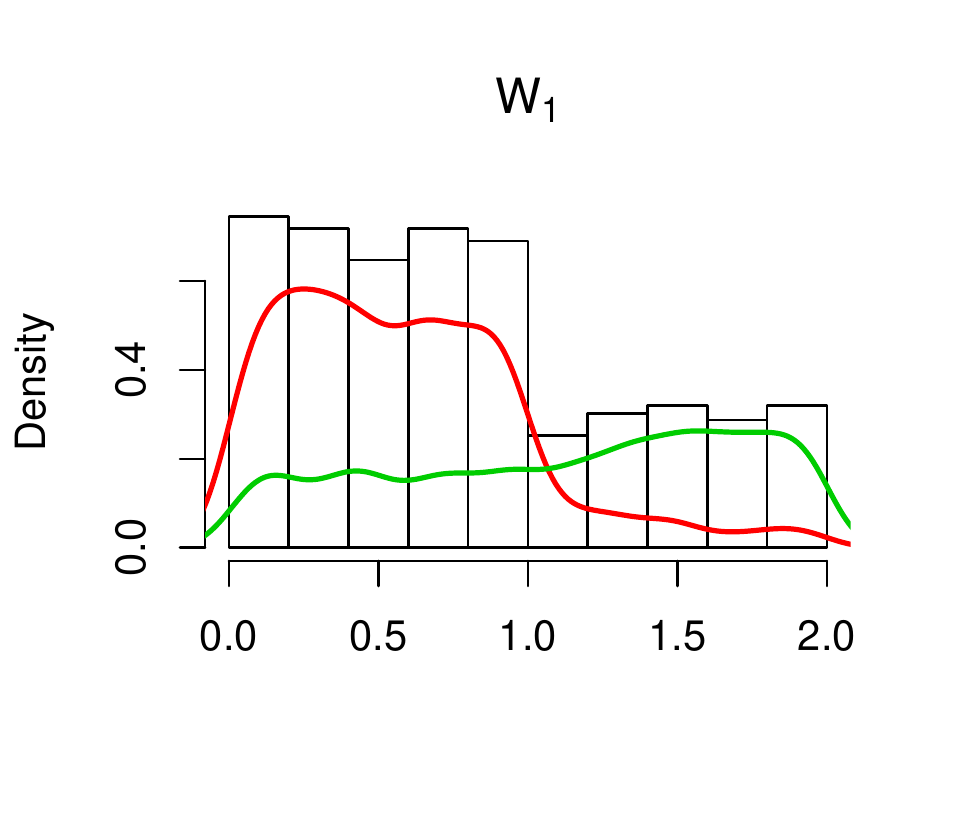}
}
\newline
\subfloat{
  \includegraphics[scale=.8]{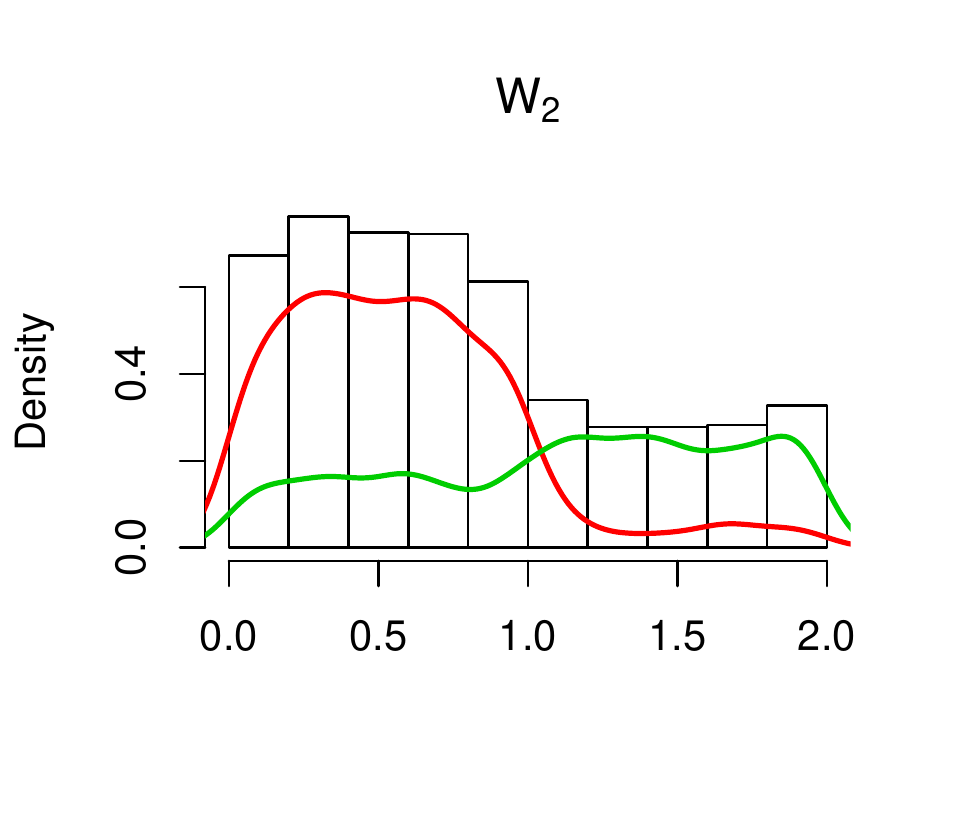}
}
\end{figure}

Rather than giving up, we recommend that the researcher try to find a subset of exogenous observations (the observations drawn from the second component) from the 2,000 observations on $(X, W_1, W_2)$. If the researcher knows or is willing to assume that $E(X_1) < E(X_2)$, she can use our method. Doing so, the researcher would:
\begin{enumerate}[noitemsep]
\item Consistently estimate the mixture model defining $(X, W_1, W_2)$ to recover the coordinate densities for each component. See Figure \ref{density}.
\item Label these components as $1$ and $2$ according to the moment condition $E(X_1) < E(X_2)$.
\item Choose observations that are drawn from the second component with probability greater than or equal to $p=.9$. Here this identifies 524 likely exogenous observations.
\end{enumerate}

These steps give a subset of observations of $X$ that are asymptotically realized from $X_2$ with probability greater than or equal to $p$ under the true mixture distribution. The researcher can then run ordinary least squares on this subset to recover $\beta$. Doing so for our realizations, gives a point estimate of 1.97, which is not significantly different from 2 at conventional levels (standard errors are calculated by bootstrapping over the entire procedure to account for the subset selection). See Column 2 of Table \ref{simulation_regs}.

\section{Application}
\label{application}
A better test of our approach is whether it can recover exogenous observations from commonly encountered economic data. To this end, we apply our method to try to find pricing experiments in retail scanner data. We consider the Dominick's scanner data maintained by the Kilts Center for Marketing. \cite{hoch1994edlp} conducted a number of pricing experiments with Dominick's in the time period covered by this data. We apply our approach to recover Hoch et al.'s ``Study 1" pricing experiments. We show that our results closely match the existing documentation on these experiments. We also provide some evidence that our results more accurately describe the observed data than this documentation.

We allow ourselves few liberties in applying our method to the Dominick's data, and so do not present a full description of Hoch et al. here. The limited background knowledge that we permit ourselves is that the authors experimented with how Dominick's grocery stores priced some products in select categories for several weeks in 1992 and 1993.\footnote{We consider the categories Analgesics, Canned Soups, Cereals, Cheese, Dish Detergent, Front End Candies, Frozen Entrees, Snack Crackers, Soft Drinks, and Toothpaste. From the 19 categories used in the paper, we omit those where we do not know the true store-category treatment levels and experiment time frame.} In their ``Study 1" pricing experiments, Hoch et al assigned each store-category pair to one of three treatment levels: ``Control," ``Hi-Lo," or ``EDLP." Stores assigned to ``Control" for a category kept pricing products following the chain's standard procedure, stores assigned to ``Hi-Lo" for a category raised prices on the specified products, and stores assigned to ``EDLP" for a category lowered prices on the specified products. We also permit ourselves to know and make use of the fact that  Dominick's split their 87 stores into 16 socioeconomic zones. Given this background, we hope to recover all store-week-category treatment labels.\footnote{See Table \ref{zones}. We exclude zones with 3 or fewer stores. We also exclude stores that had no product purchased from the considered category in over 15\% of the weeks observed. This sub-setting leaves us with an average of 54 stores in 6 zones per category.}

For a given category with $\mathbb{P} = \{1, \ldots, P\}$ products, Dominick's weekly prices over 1992-1993 map into our framework above. Let $X_{jit}$ be the log demeaned price of product $j \in \mathbb{P}$ from the category in store $i$ during week $t$.\footnote{Specifically, we subtract the zone-week-product mean price.} The $P$ product price vector, $(X_{1it}, \ldots, X_{Pit})$, is a three component mixture model. The components are the respective price distributions under ``Control," ``Hi-Lo," and ``EDLP" pricing. As mentioned above, our approach immediately extends to such three component mixtures.

To implement our method, we assume that the log demeaned prices of the products in the category at store $i$ in week $t$ are independent of each other conditional on the treatment that the store is assigned in that week. Three such prices, say $(X_{1it}, X_{2it}, X_{3it})$, then take the place of $(X, W_1, W_2)$ in the exposition above. Here, $X_{2it}$ and $X_{3it}$ do double duty. $X_{2it}$ and $X_{3it}$ allow us to disentangle the mixture distribution's component weights from its coordinate densities and are also important variables for the regression in their own right. We use $P > 3$ products because this application features a mixture model with three components. \cite{hall2005nonparametric} prove that four coordinates are necessary for identification of three component mixture models.\footnote{ \cite{hall2005nonparametric} establish that the condition $2^r-1 \ge mr + 1$ for $r$ the number of coordinates and $m$ the number of components is nessecary for nonparametric identification. No sufficient condition has been established yet. See the discussion in \cite{benaglia2009like}. Despite the lack of formal identification, the results below suggest that we are nevertheless able to recover the three components.} We know that Hoch et al. conducted their experiments using only a subset of products in a category. As such, we use all products whose maximum difference in price between stores in a zone in any week is higher than 3\%. The results are robust to using all products.

Given these preliminaries, we then apply our method directly. We:
\begin{enumerate}[noitemsep]
\item Use the algorithm from \cite{benaglia2009like} to estimate a three component mixture model from all obsevations on $(X_{1it}, \ldots, X_{Pit})$ in a zone (repeating for all zones and categories).
\item Label the three components as ``Control," ``Hi-Lo," or ``EDLP" according to the moment condition $E(\sum_{j \in \mathbb{P} } X_{jit}|Hi\text{-}Lo) > E(\sum_{j \in \mathbb{P} } X_{jit}|Control) > E(\sum_{j \in \mathbb{P} } X_{jit}|EDLP)$.
\item And predict that Hoch et al. assigned a store-week-category, $(X_{1it}, \ldots, X_{Pit})$, the treatment label that has the highest probability under the estimated mixture model.
\end{enumerate}

Table \ref{accuracy} illustrates our success at recovering the documented store-week-category treatment labels for January 1992 through December 1993. A key assessment of our accuracy is the number of store-week-category observations we correctly predict the label for across the time frame that Hoch et al. document as control and treatment for their ``Study 1" experiments. Column 3 shows that our method recovers the correct documented label for over 70\% of the store-weeks on average. Hoch et al. proceeded to conduct additional pricing experiments after the end of the ``Study 1" we consider here. This additional variation provides a lower bound on our accuracy. That is, a lower bound on our accuracy is the number of store-week-category observations we correctly predict the label for across the full two years assuming that store-week-category observations not documented in ``Study 1" are ``Control." Even by this lower bound, our labels are accurate for over 40\% of the store-weeks on average. See Column 6.

\begin{figure}[H]
\caption{Average log demeaned price for Toothpaste over time in Dominick's store 48 classified into ``Control", ``Hi-Lo", and ``EDLP." The nonparametric estimation recovers the weeks documented as ``Hi-Lo" along with additional weeks wherein the store appears to have continued following ``Hi-Lo" pricing.}
\centering
\includegraphics[scale=.75]{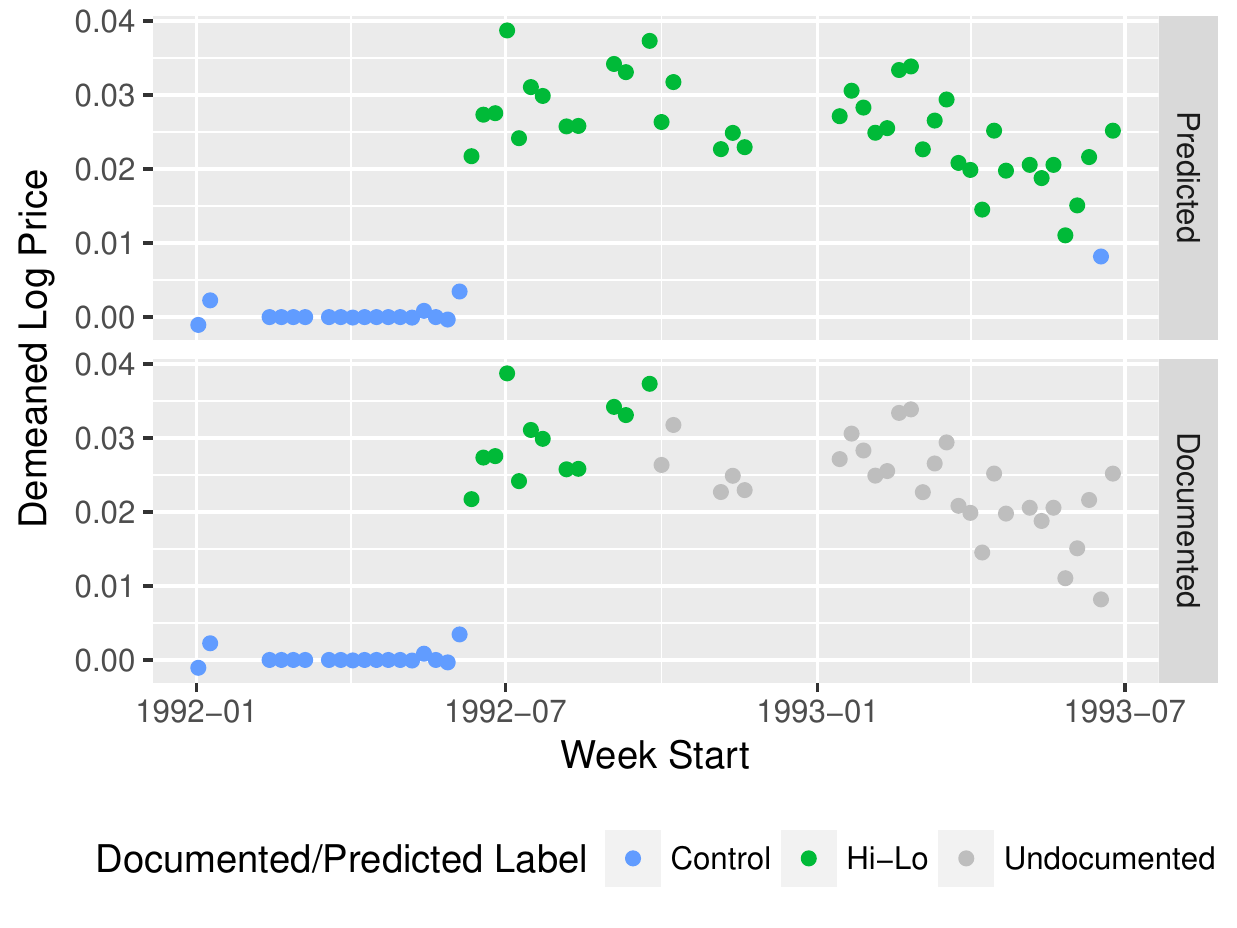}
\label{frozen}
\end{figure}

Graphically displaying the recovered exogenous variation provides additional insight into our method's success at identifying store-week-category treatment labels. Figure \ref{frozen} shows the average  log demeaned price for Toothpaste over time in Dominick's store 48 classified into ``Control", ``Hi-Lo", and ``EDLP." Panel 1 depicts the labels we predict for each week and Panel 2 depicts the documented labels for each week. Comparing the panels, it is immediate that the nonparametric estimation almost perfectly recovers both the weeks documented as ``Control" and the weeks documented as ``Hi-Lo." Further, the nonparametric estimation also appears to correctly predict that the store continued ``Hi-Lo" pricing in the months following the end of Hoch et al.'s ``Stage 1" experiment.

\begin{figure}[H]
\caption{Average log demeaned price for Canned Soups over time in Dominick's store 5 classified into ``Control", ``Hi-Lo", and ``EDLP." The nonparametric estimation correctly captures that a ``Hi-Lo" experiment present in the documentation did not happen.}
\centering
\includegraphics[scale=.75]{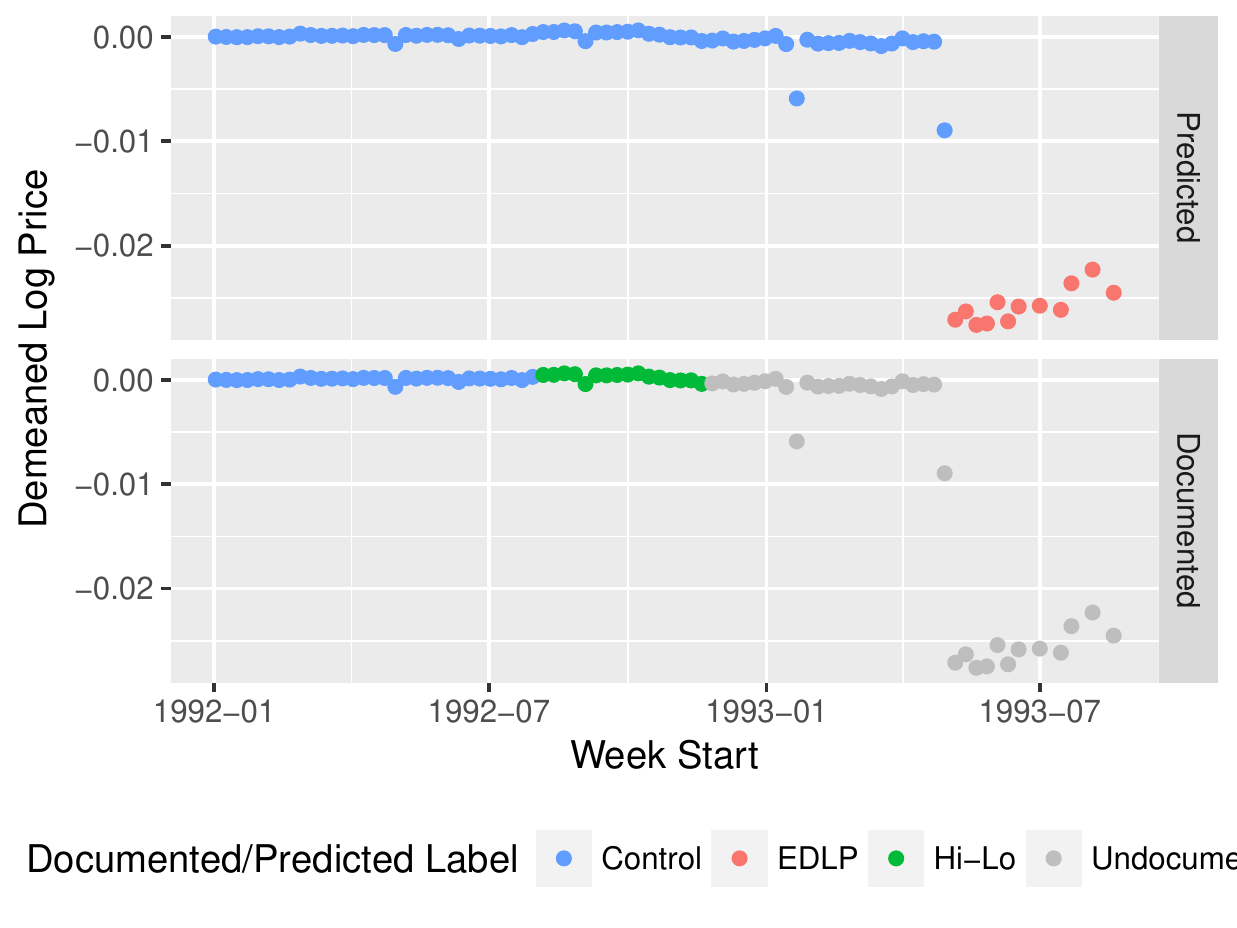}
\label{canned}
\end{figure}

\begin{figure}[H]
\caption{Average log demeaned price for Dish Detergent over time in Dominick's store 91 classified into ``Control", ``Hi-Lo", and ``EDLP." The nonparametric estimation correctly recovers that the ``Hi-Lo" experiment started later than described in the documentation.}
\centering
\includegraphics[scale=.8]{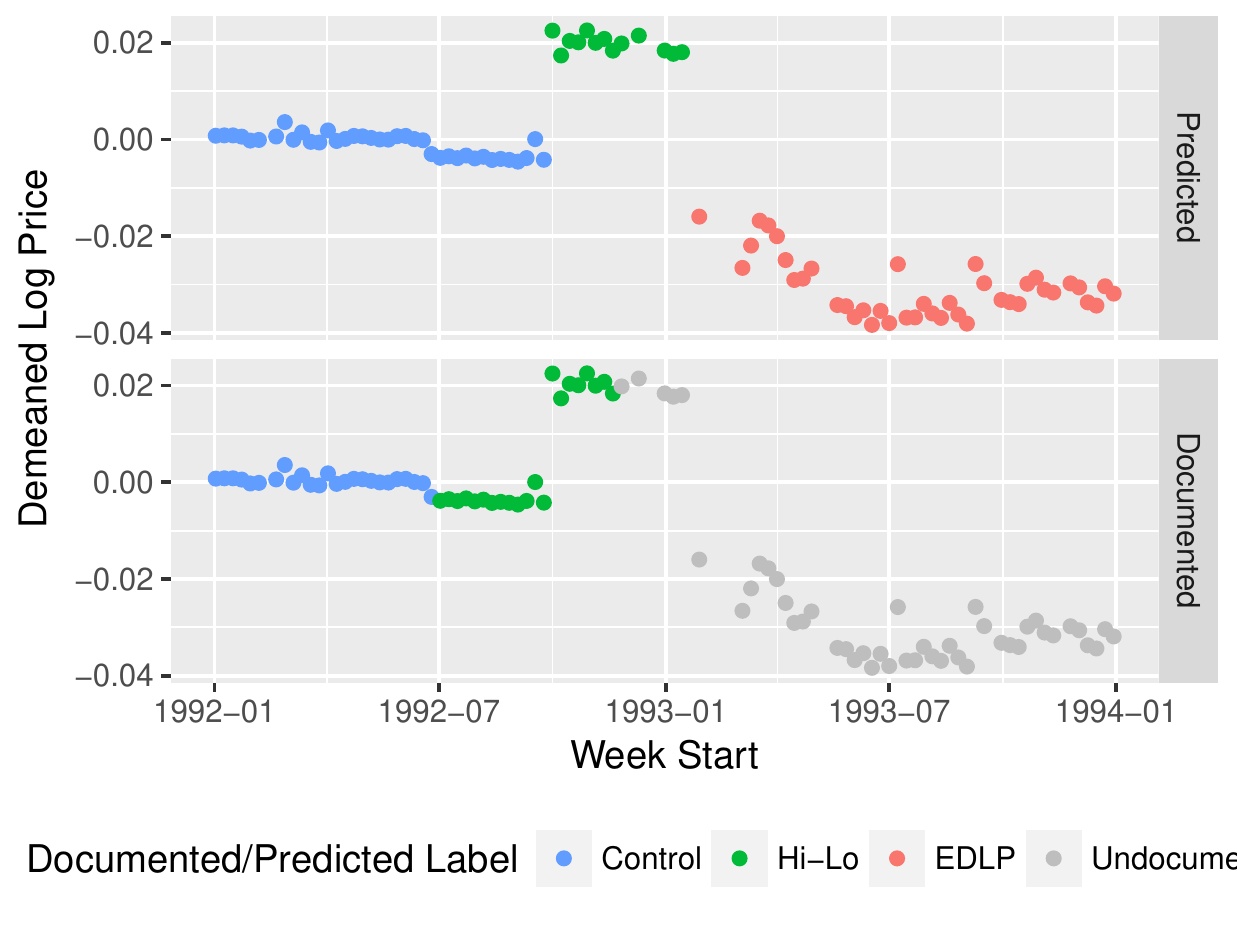}
\label{dish}
\end{figure}

\begin{figure}[H]
\caption{Average log demeaned price for Frozen Entrees over time in Dominick's store 116 classified into ``Control", ``Hi-Lo", and ``EDLP." The nonparametric estimation recovers that the documented ``Hi-Lo" experiment was not consistently implemented. Store 116 prices its frozen entrees well below the average weekly price for the zone during four weeks in the experiment period.}
\centering
\includegraphics[scale=.8]{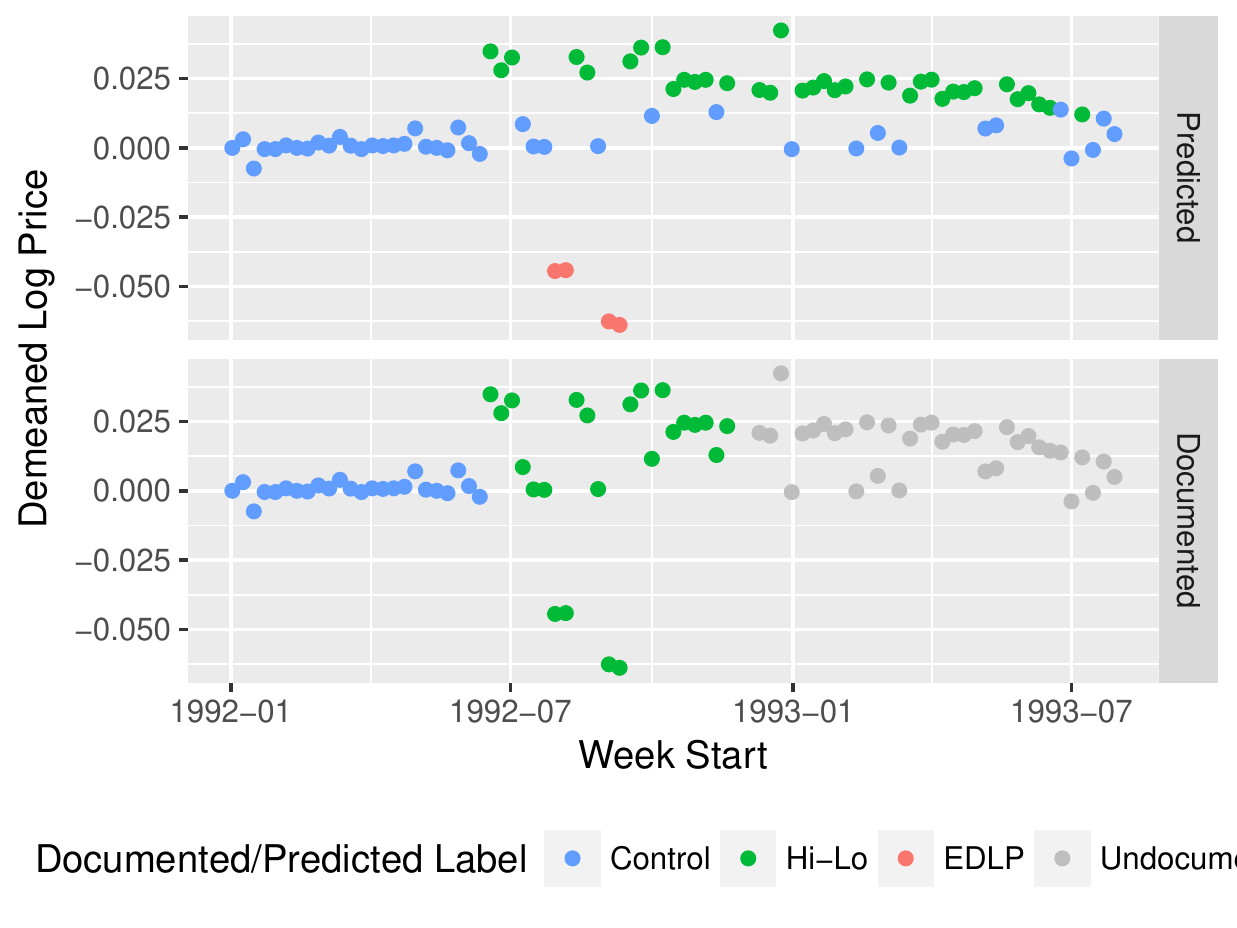}
\label{frozen2}
\end{figure}

In several cases, our predicted labels better match the observed data than the documented labels. Figure \ref{canned} shows the average  log demeaned price for Canned Soups by week in Dominick's store 5 classified into ``Control", ``Hi-Lo", and ``EDLP." Here the price data and our approach agree that the ``Hi-Lo" experiment present in the documentation did not occur. Figure \ref{dish} provides the same time series for Dish Detergent in Dominick's store 91. Here the price data and our approach imply that the ``Hi-Lo" experiment started later than described by the documentation. Finally, Figure \ref{frozen2} considers Frozen Entrees as sold by Dominick's store 116. The price data and our approach suggest that the ``Hi-Lo" experiment was not consistently implemented throughout the documented time frame. Store 116 priced its frozen entrees well below the average weekly price for the zone in four weeks. 

These discrepancies between the price data and the experiment documentation are not isolated instances. Table \ref{price} reports the price change that Hoch et al. assigned to each category for the ``Hi-Lo" experiments in Column 1 and the ``EDLP" experiments in Column 4. In Columns 2 and 5, we try to replicate these price changes using the observed prices and the documentation on the experiment time frame for the ``Hi-Lo" and ``EDLP" experiments respectively.\footnote{A caveat is that the documentation does not specify which products were used in the experiment. Here, we follow the same procedure used to pick products for the nonparametric estimation. We use all products whose maximum price difference between stores in a zone in a week is higher than 3\% during the documented experiment period. Using all products gives substantially similar results.} In both cases, the replicated price changes differ significantly from the published price changes and several even have the wrong sign. We suspect that the existing documentation reflects an original experiment design that was revised slightly before implementation and that the experiments themselves were not executed perfectly. Finally, Columns 3 and 6 present the price changes recovered for the ``Hi-Lo" and the ``EDLP" experiments using our predicted labels.\footnote{We form ``Hi-Lo" (``EDLP") price changes by averaging over store-week-category subsets consisting of six consecutive weeks labeled as control and then six consecutive weeks labeled as ``Hi-Lo" (``EDLP"). To facilitate comparison, we restrict to ``Hi-Lo" and ``EDLP" experiments recovered from the documented experiment period.} Our recovered price changes are closer to the published price changes and all of the correct sign. We conclude that the experimental variation is present and better captured by our method than by the existing documentation.

While the existing documentation is not perfect, it is still useful for benchmarking the consumer demand elasticities that we recover with our approach. To estimate the elasticity for a category, we average the store-week-product observations over the control period and over the treatment period to produce two observations for each product-store pair. We then estimate the difference-in-difference specification
$$q_{jie} = P_j + S_i + T_e + \beta p_{jie} + \epsilon_{jie}$$
The subscript $j$ refers to the product, $i$ the store, and $e$ whether the period is control or treatment. $q_{jie}$ is the log average quantity sold; $p_{jie}$ is the log average price; $P_j$, $S_i$, and $T_e$ are product, store, and treatment period fixed effects. Table \ref{elasticity}, Column 1 reports the estimated elasticities by category based on the documented experiment time frames. Column 2 reports the estimated elasticities by category based on our predicted store-week labels.\footnote{As above, we predict that a store ran an experiment when our approach labels 6 store-weeks as control followed by 6 store-weeks as experiment. We use this same time frame to form the control and treatment periods. Unlike the documentation, our approach gives that the experiments started at different times for different stores. As such, we match each experiment store to a store that we predict spent the same 12 weeks pricing only as control. We estimate the difference-in-difference specification on the dataset formed by these matched pairs.} For every category, our approach recovers an elasticity close to the elasticity estimated from the documentation.

\section{Conclusion}
\label{conclusion}
Two-stage least squares recovers exogenous variation from an endogenous regressor using an instrumental variable. Unfortunately, researchers rarely have access to a valid instrument. As such, we revisit this problem with a new approach. Our key insight is that if the regressor is a mixture of exogenous and endogenous components, then nonparametrically estimating the underlying mixture model recovers a subset of the exogenous observations. These recovered observations can then be used in a second stage to identify the parameter of economic interest.

Our method applies to the prototypical example of a widget store choosing prices at which to sell its widgets each week. These prices are endogenous in the regression of log quantity on log price because they are set simultaneously with demand. Assuming that the widget store either purposely experiments with its pricing practices for some periods of time or events arise that produce natural experiments, then there are weeks in which prices are exogenous. Our method enables the researcher to recover a subset of the experiment weeks if she knows either (A) whether experiment weeks are more or less common than non-experiment weeks or (B) how a moment of an observed variable differs between experiment and non-experiment weeks.

Our approach has promise. It recovers exogenous observations in simulation and can be used to find pricing experiments hidden in retail scanner data. That is, an researcher given observations on $(Y, X, W_1, W_2)$ satisfying Assumption \ref{one} can still recover $\beta$ without either $W_1$ or $W_2$ being valid instruments for $X$. In practice, a researcher given scanner data from a grocery store chain covering weeks in which the chain ran a pricing experiment can recover the essential details of the experiment--how each store was assigned to price products in each category each week. 

As the Hoch et al. application makes prominent, one future application of our method is in facilitating replication exercises. We believe that many older economics papers utilize experiments for which the documentation has not been fully preserved. In such cases, our approach can be used to supplement the existing documentation.

Similarly, our understanding is that many websites conduct A/B tests to improve their services and do not preserve any information on these tests. Consider a movie recommendation website that tests its personalization algorithm by showing random recommendations to randomly selected visitors. A researcher with data on which of $M$ movies were recommended to a homogenous set of users could use our method to identify a subset of the users shown random recommendations. Let $X_{ij}$ be whether movie $i$ was recommended to user $j$. Then the $M$ vector, $(X_{1j}, \ldots, X_{Mj})$, is a two component mixture model. The researcher could label the components using the assumption that the probability of being shown random recommendations is small, $\pi < 0.5$.

An immediate extension of our method exists in situations where the researcher has partial information on experiments in the data. For example, the researcher may know details of some experiments that occur within the data and wish to find additional similar experiments. Here, the known details can be used to semiparametrically estimate the underlying mixture model in place of relying on nonparametric estimation.

There are many settings where researchers currently struggle to recover exogenous variation. We hope that our approach will facilitate advances in at least some of these instances. We expect that lessons from applying nonparametric estimation techniques will be informative on additional approaches to identification in the future.

\newpage
\singlespacing

\bibliographystyle{apalike}
\bibliography{Bibliography}

\appendix
\newpage
\section{Proof of Theorem \ref{first}}
\label{proof}
Let $\chi(p)$ be a subset of the set of observations of $(Y_t, X_t)$ where $X_t$ is drawn from $X_2$ with probability greater than or equal to a threshold $p$ under the true mixture model. Assume (1) $|\chi(p)|$ approaches infinity as $T$ approaches infinity and (2) the $Cov(X,\epsilon|X\in\chi(p))$ and $Var(X|X\in\chi(p))$ are finite for all $p\in[0,1]$

By (1)
\begin{align*}
\hat{\beta}^{OLS}_{\chi(p)} &\underset{T\to\infty}{\overset{p}{\to}} \frac{Cov(Y,X|X\in\chi(p)) }{Var(X|X\in\chi(p))}
\end{align*}

\noindent Applying the law of total variance to the denominator, which is valid under assumption (2), gives
\begin{align*}
Var(X|X\in\chi(p)) &= E[Var(X|1_{Exp},X\in \chi(p))|X\in \chi(p)] + Var(E[X|1_{Exp}, X\in \chi(p)]| X\in \chi(p))
\\ &= (1-p) Var(X|1_{Exp}=0,X\in \chi(p)) + p Var(X|1_{Exp}=1,X\in \chi(p)) \\ &+ Var(E[X|1_{Exp}, X\in \chi(p)]|X\in \chi(p))
\\ &= (1-p) Var(X_1|X_1 \in \chi(p)) \\ &+ p Var(X_2|X_2 \in \chi(p)) + Var(E[X|1_{Exp},X\in \chi(p)]|X\in \chi(p))
\end{align*}

\noindent Applying the law of total variance to the numerator gives
\begin{align*}
Cov(Y,X|X\in\chi(p)) &= E[Cov(Y,X|1_{Exp},X\in \chi(p))|X\in \chi(p)] \\ &+ Cov(E[Y|1_{Exp}, X\in \chi(p)], E[X|1_{Exp}, X\in \chi(p)]|X\in \chi(p))
\end{align*}

\noindent Then
\begin{align*}
E[Cov(Y,X|1_{Exp},X\in \chi(p))|X\in \chi(p)] &= (1-p) Cov(Y,X|1_{Exp}=0,X\in \chi(p)) \\ &+ p Cov(Y,X|1_{Exp}=1,X\in \chi(p)) 
\\ &= (1-p) Cov(Y_1,X_1|X_1 \in \chi(p)) + p Cov(Y_2,X_2|X_2\in \chi(p))
\end{align*}
where $Y_1$ indicates that the data-generating process for $Y$ is in terms of $X_1$ only and similar for $Y_2$

As $p\to1$, 
\begin{align*}
(1-p) Var(X_1|X_1 \in \chi(p)) + p Var(X_2|X_2 \in \chi(p)) &\to Var(X_2|X_2 \in \chi(1)) \\
Var(E[X|1_{Exp},X\in \chi(p)]|X\in \chi(p)) &\to 0 \\
(1-p) Cov(Y_1,X_1|X_1 \in \chi(p)) + p Cov(Y_2,X_2|X_2\in \chi(p)) &\to Cov(Y_2,X_2|X_2 \in \chi(1)) \\
Cov(E[Y|1_{Exp}, X\in \chi(p)], E[X|1_{Exp}, X\in \chi(p)]|X\in \chi(p)) &\to 0
\end{align*}

So
\begin{align*}
\hat{\beta}^{OLS}_{\chi(p)} & \underset{T\to\infty}{\overset{p}{\to}} \frac{Cov(Y,X|X\in\chi(p)) }{Var(X|X\in\chi(p))}\\
& \underset{p\to 1}{\to} \frac{Cov(Y_2,X_2|X_2 \in \chi(1))}{Var(X_2|X_2 \in \chi(1))} \\
& = \frac{Cov(\alpha + \beta X_2,X_2|X_2 \in \chi(1))}{Var(X_2|X_2 \in \chi(1))} \\ &= \beta
\end{align*}

\newpage
\section{Figures}

\newpage
\section{Tables}

\begin{table}[H]
\singlespacing
\caption{Regression results from 2,000 observations of the random vector $(Y, X, \epsilon, W_1, W_2)$ generated as described in Section \ref{simulation}. Column (1) displays the coefficient estimated using OLS on the full sample. Column (2) displays the coefficient estimated from observations that are 90\% or more likely to have been drawn from $X_2$. Standard errors are calculated by bootstrapping over the entire procedure to account for the subset selection.} 
\label{simulation_regs}
\centering 
\begin{tabular}{@{\extracolsep{5pt}}lcc} 
\\[-1.8ex]\hline 
\hline \\[-1.8ex] 
 & \multicolumn{2}{c}{\textit{Dependent variable:}} \\ 
\cline{2-3} 
\\[-1.8ex] & $Y$ & $Y_p$ \\ 
\\[-1.8ex] & (1) & (2)\\ 
\hline \\[-1.8ex] 
 X & 2.117$^{***}$ &  \\ 
  & (0.024) &  \\ 
  & & \\ 
 $X_p$ &  & 1.968$^{***}$ \\ 
  &  & (0.056) \\ 
  & & \\ 
\hline \\[-1.8ex] 
Observations & 2,000 & 524 \\ 
R$^{2}$ & 0.800 & 0.749 \\ 
\hline 
\hline \\[-1.8ex] 
\textit{Note:}  & \multicolumn{2}{r}{$^{*}$p$<$0.1; $^{**}$p$<$0.05; $^{***}$p$<$0.01} \\ 
\end{tabular} 
\end{table}

\begin{table}[H]
\caption{Dominick's split their 87 stores into 16 socioeconomic zones. We use all zones that had at least one store assigned to a treatment and another store assigned to control.}
\centering
\include{robustness/zone_selection}
\label{zones}
\end{table}

\begin{table}[H]
\singlespacing
\caption{Assessment of our accuracy in recovering the documented store-week-category treatment labels for Hoch et al.'s ``Stage 1" pricing experiments. Columns 1-3 consider the documented control and treatment time frames for each category. Columns 4-6 consider the entirety of 1992 and 1993 wherein the majority of weeks are undocumented and assumed to be ``Control." By and large, we are able to correctly recover the store-week-category treatment labels.}
\centering
\input{robustness/diff_period/max_zone_week/accuracy.tex}
\label{accuracy}
\end{table}

\begin{table}[H]
\singlespacing
\caption{Comparison of Hoch et al.'s assigned price changes to the constructed price changes by category. Columns 1 and 4 report the price changes that Hoch et al. assigned to each category for the ``Hi-Lo" experiments and the ``EDLP" experiments respectively. Columns 2 and 5 present the replicated price changes for the ``Hi-Lo" and ``EDLP" experiments. Columns 3 and 6 present the price changes recovered for the ``Hi-Lo" and the ``EDLP" experiments using our predicted labels. We believe that the documented experiments were not executed perfectly and that our predicted labels better reflect the actual experimental variation.}
\centering
\input{robustness/diff_period/max_zone_week/price_change_robust.tex}
\label{price}
\end{table}




\begin{table}[H]
\caption{Consumer demand elasticities estimated from Hoch et al.'s ``Stage 1" pricing experiments using a difference-in-difference specification. Column 1 reports the estimates based on the documented store-week labels. Column 2 reports the estimates based on our predicted store-week labels. Standard errors are clustered by store. The elasticities recovered by our approach are very similar to the replicated elasticities.}
\singlespacing
\centering
\input{robustness/diff_period/max_zone_week/did_reg.tex}
\label{elasticity}
\end{table}



\end{document}

%% file: robustness/zone_selection.tex
\begin{tabular}{rr}
\toprule
Zone ID & Number of Stores\\
\midrule
\addlinespace[0.3em]
\multicolumn{2}{l}{\textbf{Selected}}\\
\hspace{1em}1 & 15\\
\hspace{1em}2 & 28\\
\hspace{1em}5 & 4\\
\hspace{1em}6 & 8\\
\hspace{1em}7 & 5\\
\hspace{1em}12 & 10\\
\addlinespace[0.3em]
\multicolumn{2}{l}{\textbf{Excluded}}\\
\hline
\hspace{1em}3 & 1\\
\hspace{1em}4 & 1\\
\hspace{1em}8 & 3\\
\hspace{1em}10 & 3\\
\hspace{1em}11 & 2\\
\hspace{1em}13 & 1\\
\hspace{1em}14 & 1\\
\hspace{1em}15 & 3\\
\hspace{1em}16 & 1\\
\hspace{1em}NA & 1\\
\bottomrule
\end{tabular}

%% file: robustness/diff_period/max_zone_week/accuracy.tex
\begin{tabular}{l|ccc|ccc}
\toprule
\multicolumn{1}{c}{\bfseries  } & \multicolumn{3}{c}{\bfseries Within Experiment Period} & \multicolumn{3}{c}{\bfseries All Data} \\
\cmidrule(l{2pt}r{2pt}){2-4} \cmidrule(l{2pt}r{2pt}){5-7}
Category & \# Stores-Weeks & \# Correct & Accuracy & \# Store-Weeks & \# Correct & Accuracy\\
\midrule
Analgesics & 448 & 410 & 0.915 & 3328 & 1312 & 0.394\\
Canned Soups & 1040 & 461 & 0.443 & 5005 & 2448 & 0.489\\
Cereals & 910 & 835 & 0.918 & 5395 & 2757 & 0.511\\
Cheese & 858 & 312 & 0.364 & 4950 & 1683 & 0.340\\
Dish Detergent & 987 & 573 & 0.581 & 4559 & 1774 & 0.389\\
Front End Candies & 840 & 659 & 0.785 & 4704 & 1836 & 0.390\\
Frozen Entrees & 1334 & 783 & 0.587 & 4756 & 2108 & 0.443\\
Snack Crackers & 944 & 875 & 0.927 & 5133 & 2575 & 0.502\\
Soft Drinks & 800 & 741 & 0.926 & 3650 & 2130 & 0.584\\
Toothpaste & 592 & 494 & 0.834 & 3441 & 1327 & 0.386\\
\bottomrule
\end{tabular}

%% file: robustness/diff_period/max_zone_week/price_change_robust.tex
\begin{tabular}{lrrrrrr}
\toprule
\multicolumn{1}{c}{ } & \multicolumn{3}{c}{Hi-Lo} & \multicolumn{3}{c}{EDLP} \\
\cmidrule(l{2pt}r{2pt}){2-4} \cmidrule(l{2pt}r{2pt}){5-7}
 & Published & Replicated & Recovered & Published & Replicated & Recovered\\
\midrule
Analgesics & 10 & 2.419 & 3.014 & -10 & -3.031 & -2.683\\
Canned Soups & 14 & -0.100 & 2.353 & NA & NA & -4.798\\
Cereals & 10 & 3.978 & 5.204 & -10 & -5.565 & -6.252\\
Cheese & 8 & -0.026 & 1.881 & -8 & -0.027 & NA\\
Dish Detergent & 6 & 2.963 & 3.526 & -6 & -1.112 & -3.560\\
Front End Candies & 13 & 0.811 & 4.779 & -13 & -0.693 & -5.837\\
Frozen Entrees & 11 & 0.066 & 4.819 & -11 & -0.056 & NA\\
Snack Crackers & 10 & 1.208 & 3.367 & -10 & -1.387 & -2.736\\
Toothpaste & 7 & 3.778 & 3.855 & -7 & -4.374 & -4.097\\
\bottomrule
\end{tabular}

%% file: robustness/diff_period/max_zone_week/did_reg.tex
\begin{tabular}{lll}
\toprule
Category & Replicated & Recovered\\
\midrule
Analgesics & -1.783*** & -1.264***\\
 & (0.148) & (0.349)\\
Canned Soups & -1.186*** & -2.003***\\
 & (0.23) & (0.095)\\
Cereals & -2.723*** & -2.782***\\
 & (0.151) & (0.138)\\
Cheese & -1.45*** & -1.888***\\
 & (0.043) & (0.078)\\
Dish Detergent & -4.655*** & -4.559***\\
 & (0.154) & (0.209)\\
Front End Candies & -1.919*** & -1.119***\\
 & (0.077) & (0.408)\\
Frozen Entrees & -3.507*** & -3.728***\\
 & (0.078) & (0.133)\\
Snack Crackers & -3.143*** & -2.414***\\
 & (0.079) & (0.188)\\
Toothpaste & -5.043*** & -1.939***\\
 & (0.3) & (0.152)\\
\bottomrule
\end{tabular}